# A-DisETrac Advanced Analytic Dashboard for Distributed Eye Tracking

Yasasi Abeysinghe, Old Dominion University, USA*
https://orcid.org/0000-0002-5114-9732

Bhanuka Mahanama, Old Dominion University, USA

Gavindya Jayawardena, Old Dominion University, USA

Yasith Jayawardana, Old Dominion University, USA

Mohan Sunkara, Old Dominion University, USA
https://orcid.org/0000-0002-6970-0203

Andrew T. Duchowski, Clemson University, USA

Vikas Ashok, Old Dominion University, USA

Sampath Jayarathna, Old Dominion University, USA

**ABSTRACT**

Understanding how individuals focus and perform visual searches during collaborative tasks can help improve user engagement. Eye tracking measures provide informative cues for such understanding. This article presents A-DisETrac, an advanced analytic dashboard for distributed eye tracking. It uses off-the-shelf eye trackers to monitor multiple users in parallel, compute both traditional and advanced gaze measures in real-time, and display them on an interactive dashboard. Using two pilot studies, the system was evaluated in terms of user experience and utility, and compared with existing work. Moreover, the system was used to study how advanced gaze measures such as ambient-focal coefficient K and real-time index of pupillary activity relate to collaborative behavior. It was observed that the time a group takes to complete a puzzle is related to the ambient visual scanning behavior quantified and groups that spent more time had more scanning behavior. User experience questionnaire results suggest that their dashboard provides a comparatively good user experience.

## 1 INTRODUCTION

With the spread of COVID-19, many organizations and individuals resorted to using online platforms to interact and collaborate amidst geographic restrictions. The education industry, for instance, witnessed a large uptick in virtual learning, video conferencing, and remote collaborative activities. However, the nature of remote interaction makes individuals more susceptible to mindwandering and task disengagement (Cotton et al. 2023). Being able to monitor the visual attention and mental effort of each interacting party in parallel, possibly in real-time, may help in navigating through such barriers. Moreover, such information may help in developing tools, communication strategies, and work processes that adapt to one's cognitive and visual capacity. Furthermore, realtime gaze analytics can be shared analogous to methods like shared gaze (Zhao et al. 2023), where individuals see each other's gaze position on AR/VR (Blattgerste et al. 2018). Overall, the advancements in eye-tracking technology provide a strong foundation to both assess and improve the quality of remote interaction (D'Angelo et al. 2021; Jermann et al. 2011).

Traditional eye-tracking measures such as fixations, saccades, micro-saccades, and pupil diameter, and advanced eye-tracking measures, such as focal/ambient coefficient and low/high index of pupillary activity, have been widely utilized to study human visual attention (Jayawardena et al. 2020b; Krejtz et al. 2014, 2015, 2016) and cognitive load (Duchowski et al. 2018, 2020; Jayawardena et al. 2022b; Krejtz et al. 2018). Despite wide adoption, these measures are geared towards single-user studies and are thus challenging to scale for multi-user studies. To elaborate, since eye trackers are designed to capture eye movements of one individual at a time, eye-tracking studies are often carried out as single-user experiments (Jayawardena et al. 2021b; Mahanama et al. 2022c; Michalek et al. 2019; Senarath et al. 2022) in isolated environments (Mahanama 2022a, 2021). Moreover, these measures only capture individual-level behaviors and do not account for inter-individual interactions. As a result, the development of advanced measures geared toward multi-user environments plays a crucial role in analyzing eye-tracking data in our natural collaborative environments.

Recent advancements in multi-user eye-tracking (Pathirana et al. 2022), such as distributed eyetracking (Mahanama et al. 2023), have enabled the real-time measurement of user collaboration during online activities (Garcia et al. 2003; Guo et al. 2013; Langner et al. 2022; Mahanama et al. 2023). These methods primarily use traditional measures to estimate visual attention and cognitive assessment. Unlike advanced measures, they do not leverage the underlying associations between pupil and gaze responses. Therefore, we design our system to support real-time computation of both traditional (fixation duration, saccade duration, and saccade amplitude) and advanced gaze measures, namely, Ambient/Focal Attention with Coefficient $\mathcal{K}$ (Krejtz et al. 2016) and Real-time Index Pupillary Activity (RIPA) (Jayawardena et al. 2022b). The readers are referred to (Mahanama et al. 2022b) for a comprehensive review of the various existing gaze measures including advanced measures. We improve the reproducibility of our system by allowing users to restream eye-tracking data, and thereby mimic real-time data acquisition. Our key contributions are as follows:

1. We propose a distributed multi-user eye-tracking system that supports both advanced and traditional gaze measures.
2. We visualize both advanced and traditional measures in an interactive dashboard with restreaming support.
3. We demonstrate the utility of our system via two distributed eye-tracking user studies.

A short demo of our system is available at https://youtu.be/20LzU9NmF4o.

## 2 RELATED WORK

### 2.1 Eye Tracking Visualization

Eye tracking technology allows researchers to gain insights into human behaviors, attention, and mental effort using their eye movements. While statistical analysis of eye-tracking data provides quantitative results to support or reject hypotheses, eye-tracking visualization provides a way for exploratory and qualitative analysis of data (Blascheck et al. 2017). Through eye-tracking visualizations, researchers can spot trends and unexpected gaze patterns that may not be immediately apparent through statistical analysis alone and find new hypotheses to investigate.

Existing eye movement analysis tools primarily visualize data through heat maps, gaze plots, AOI plots, dynamic pupil diameter/gaze plots, and main sequence plots. Among these, heat maps (attention maps), perception maps (gaze opacity maps), and gaze plots (scanpaths) are often used to visualize spatio-temporal relationships in gaze data (Drusch et al. 2014; Špakov et al. 2007). Likewise, statistical graphs, such as mean fixation duration, mean saccade amplitude, and mean saccade velocity, are often used to summarize gaze data across time (Berg et al. 2009; Convertino et al. 2003; Smith et al. 2013). Both prior work (Duchowski et al. 2012) and commercial applications (Tobii[1], Tableau[2]) employ heatmaps, perception maps, gaze plots, and object-based color coding to visualize eye movement data. However, these visualizations are generated post-data collection and thus computed offline. Real-time visualizations, on the other hand, allow monitoring of temporal trends and recurring patterns of user behavior during exploratory data analysis. For instance, tools such as DisETrac (Mahanama et al. 2023) generate real-time visualizations of gaze data, as well as real-time graphs of traditional gaze measures. Yet, they only visualize spatio-temporal data and traditional gaze measures, and lack support for advanced gaze measures such as Ambient/Focal Coefficient $\mathcal{K}$ and RIPA. Without these, its difficult to study how one's visual attention and cognitive load changes during a task.

### 2.2 Collaborative Games

In multiplayer games, each player performs a collaborative or competitive task, with their individual decisions have a positive or negative impact on their winning chances. Here, each player can process information differently, meaning they could exhibit different cognitive loads and attention patterns. Several studies have been conducted to understand how collaborative multiplayer gaming affects the players' cognition and attention during gameplay. In particular, they used questionnaire-based methods (Alharthi et al. 2018, 2021) and qualitative methods (Ang et al. 2007) to assess the mental workload during gameplay. While such methods capture the player's perceived mental effort, difficulty, and workload, those measures are highly subjective. Thus, analyzing objective measures such as EEG (Johnson et al. 2015; Wikström et al. 2022), and eye movement data during gameplay could be useful to study the cognition and attention of each player.

### 2.3 Multi-User Eye Tracking

Multi-user eye tracking studies can be categorized into two categories: *time-sharing* and *space sharing* (Pathirana et al. 2022). In time-sharing methods, eye-tracking data is captured from each user across different time windows. In contrast, space-sharing methods capture eye-tracking data from multiple users simultaneously. Space-sharing is more prominent in the literature as it allows to study phenomena such as joint visual attention and joint mental effort. However, since traditional eye-tracking setups can only track one person at a time (Mahanama 2022a), space-sharing methods require a dedicated eye tracker per user (Cheng et al. 2022a; D'Angelo et al. 2016a, 2018; Zhang et al. 2017). While space-sharing is straightforward for physically co-located users, it is challenging to execute in a distributed setting. Here, data collection and real-time analytics computation are affected by network latency, bandwidth, and traffic. In our previous work (Mahanama et al. 2023), we proposed a distributed eye-tracking framework, DisETrac, and demonstrated its utility through a pilot study. In that work, we conducted a two-player puzzle-solving task, computed traditional eye-tracking measures in realtime, and reported the network latency observed. However, we did not evaluate how users perceive the DisETrac system and the gaze measures provided therein.

### 2.4 Gaze Measures for Remote Collaborative Tasks

Shared gaze (D'Angelo et al. 2021) is a technique to view each other's gaze information during remote, collaborative tasks, and thereby communicate via non-verbal cues. Its effects have been studied on several collaborative tasks, including learning (Sharma et al. 2016; Stein et al. 2004), programming (Cheng et al. 2022b; D'Angelo et al. 2017; Stein et al. 2004), co-writing (Kütt et al. 2019), meeting (Langner et al. 2022), puzzle solving (D'Angelo et al. 2016b), game playing (Špakov et al. 2019; Zhao et al. 2023), and visual search (Brennan et al. 2008; Neider et al. 2010). While such visualization is possible for distributed multi-user eye-tracking setups, real-time gaze analytics visualization during collaborative tasks remains largely unexplored. In our prior work (Abeysinghe et al. 2023; Mahanama et al. 2023), we introduced an analytics dashboard that provides real-time visualizations of individual and aggregate measures in the distributed multi-user eye-tracking system.

Eye-tracking measures such as joint visual attention have been used to study collaborative behavior during interactive tasks (Mahanama et al. 2023; Schneider et al. 2018; Sharma et al. 2014). Existing studies have often tailored these measures to task-specific conditions (D'Angelo et al. 2021). For instance, DisETrac (Mahanama et al. 2023) computes joint visual attention from the distance between gaze position centroids of each user. With-me-ness (Sharma et al. 2014) computes joint visual attention by aggregating the entry time, first fixation duration, and the number of revisits. These studies also computed pupillary measures and traditional gaze measures such as fixations and saccade-related metrics (Mahanama et al. 2023; Sharma et al. 2016).

In this work, we extend DisETrac to support advanced gaze measures for collaborative interactions, in addition to traditional gaze measures. In particular, we integrate dynamic Ambient/Focal attention with coefficient $\mathcal{K}$ and RIPA with the gaze analytics dashboard. Using these measures, we study the visual search behavior and cognitive load during collaborative tasks.

## 3 METHODOLOGY

### 3.1 Real-Time Eye-Tracking Setup

We use the distributed eye-tracking setup proposed by DisETrac (Mahanama et al. 2023) for our experiments, comprising two main components for eye-tracking; (1) data acquisition and transmission, (2) aggregation and visualization. Similar to the DisETrac study, we sample data from common off-the-shelf eye trackers using the vendor API/SDK. Then we transmit data to an MQTT broker through a public network. MQTT broker is a message server that facilitates communication between publisher and subscriber clients.

In our setup, we acquire the gaze position of each user on the screen $(x,y)$ and the pupil dilation of each user, along with confidence estimates as determined by the vendor software. Prior to data transmission, we append an originating timestamp and a sequence number to recover the temporal order at the processing end. Moreover, we periodically perform clock synchronization using Network Time Protocol (NTP).

### 3.2 Data Restreaming Setup

We integrated StreamingHub (Jayawardana et al. 2022) to restream data from earlier experiments. In particular, given a time-series recording $R(i) = (t_i, d_i)$ where $t$ is the timestamp, $d$ is the measured quantity at $t$, and $i = 0…N$ is an indexing parameter, StreamingHub periodically emits data points with $t_i \leq (t_0 + \tau)$, where $\tau$ is the time elapsed since the $t_0$ is restreamed. Internally, StreamingHub handles file-system access and data loading and provides a storage-agnostic Python API to restream data. In our setup, we call this API to restream data from earlier experiments and direct this data onto an MQTT broker through a public network.

In both real-time setup and data re-streaming setup, at the processing end, we subscribe to the eye-tracking data streams of the MQTT broker and use them to compute eye-tracking measures. We utilize user identifier information to distinguish and compute eye-tracking measures for each user, which we then use to compute aggregate measures. Finally, we present the data to a proctor through an interactive dashboard. The overall architecture of our setup is shown in Figure 1.

**Figure 1. Architecture of the proposed distributed eye tracking system for visual attention and cognitive load. In a real-time distributed eye-tracking system, common off-the-shelf eye-trackers are used to collect data from multiple users. In the data restreaming setup, StreamingHub is called to restream data from existing data/experiments and transmit it to the MQTT broker. The realtime traditional positional and advanced gaze measures are calculated by passing the data through RAEMAP (Jayawardena et al. 2020a).**

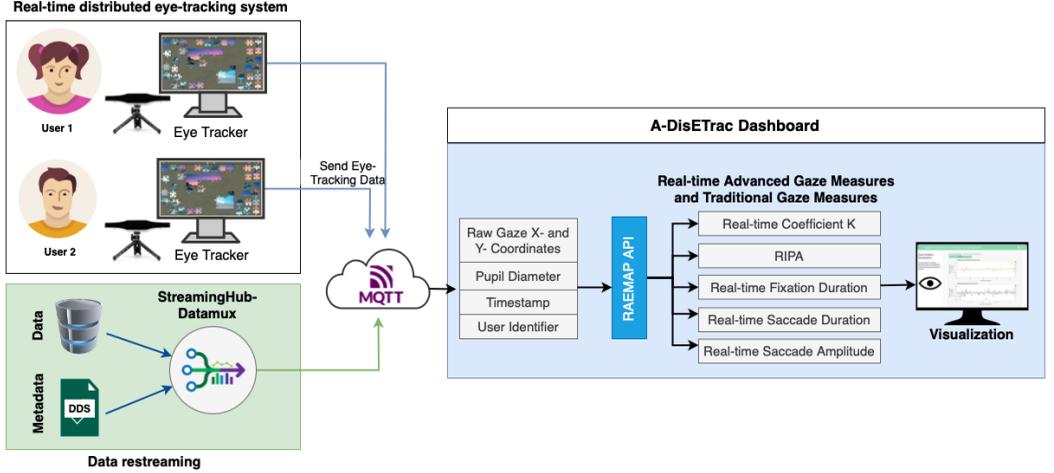

## 3.3 Real-time Gaze Measures

For our computations, we use Real-Time Advanced Eye Movements Analysis Pipeline (RAEMAP) (Jayawardena 2022a, 2020a, 2021a), an eye movement processing library to compute real-time gaze measures in two steps generating, (1) traditional positional gaze measures for each user, and (2) advanced gaze measures for each user and the group.

### 3.3.1 Coefficient $\mathcal{K}$

For each user, we order incoming data by timestamp, and compute a sequence of $\mathcal{K}$ values using a sliding window. Inside each window, we first identify fixations, i,e., periods where the gaze remains stationary, and saccades, i.e., periods where the gaze shifts rapidly (Mahanama et al. 2022b). Second, we compute the duration ($d$) of each fixation, and the saccade amplitude ($a$) between consecutive fixations. Third, we use these $d, a$ values to compute a windowed Ambient/Focal Attention Coefficient $^w\mathcal{K}$, which indicates visual search behavior within each window. Unlike $\mathcal{K}$, which uses global statistics of $d$ and $a$, $^w\mathcal{K}$ uses per-window statistics of $d$ and $a$ as a proxy for global statistics. For a time window $w$, $^w\mathcal{K}$ is defined as,

$$^w\mathcal{K} = \frac{1}{|w|} \sum_{i \in w} {}^w\mathcal{K}_i ; {}^w\mathcal{K}_i = \left( \frac{d_i - \mu_{w,d}}{\sigma_{w,d}} \right) - \left( \frac{a_{i+1} - \mu_{w,a}}{\sigma_{w,a}} \right) \qquad (1)$$

where $\mu_{w,d}$, $\sigma_{w,d}$, $\mu_{w,a}$, and $\sigma_{w,a}$ are the mean and standard deviation of $d$ and $a$ within a window $w$, $a_{i+1}$ is the saccade amplitude after the $i^{th}$ fixation, and $d_i$ is the duration of the $i^{th}$ fixation. Unlike $\mathcal{K}$, $^w\mathcal{K}$ is a windowed operation, meaning it only needs data within a window $w$ to function, and can be rerun on different time windows. However, we note that our method takes biased estimates of global statistics in computing $^w\mathcal{K}$. For experiments beyond the window $w$, we compute an aggregate coefficient $^W\mathcal{K}$ by averaging $^w\mathcal{K}$ across all $w$.

$$^W\mathcal{K} = \frac{1}{|W|} \sum_{w \in W} {}^w\mathcal{K} \tag{2}$$

For a multi-user environment, we extend $^w\mathcal{K}$ and $^W\mathcal{K}$ with group coefficients, which are computed by averaging across all users either in a specific time window ($^{Uw}\mathcal{K}$) or across the entire experiment ($^{UW}\mathcal{K}$).

$$^{Uw}\mathcal{K} = \frac{1}{|U|} \sum_{u \in U} {}^w\mathcal{K}; {}^{UW}\mathcal{K} = \frac{1}{|U|} \sum_{u \in U} {}^W\mathcal{K} \tag{3}$$

Our study uses $^{Uw}\mathcal{K}$ for real-time visualization and $^{UW}\mathcal{K}$ to compare performance across groups. We note that our modifications to $\mathcal{K}$ do not affect the interpretation of its values, meaning that positive values of $^w\mathcal{K}$, $^W\mathcal{K}$, $^{UW}\mathcal{K}$, and $^{Uw}\mathcal{K}$ indicate ambient visual scanning and negative values indicate focal processing.

### 3.3.2 Real-Time Index of Pupillary Activity (RIPA)

We also compute the RIPA (Jayawardena et al. 2022b) which is an indicator of cognitive load. RIPA is a real-time measure of pupil diameter oscillation that is computed via the ratio between two finite-impulse response filters. Savitzky et al. 1964 and Golay formulated data smoothing and derivative computation as least-square approximation problems. They equated both processes to convolving with a kernel whose coefficients are computed in closed form for a given window size and polynomial degree. In particular, given a pupil diameter signal $x[t]$, a window size of $2m + 1$, and a polynomial order of $n$, its smoothed approximation $p[t; A]$ is given by,

$$p[t; A] = \sum_{k=0}^{n} a_k t^k = a_0 + a_1 t^1 + \ldots + a_n t^n \tag{4}$$

and its derivative $p'_{[t; A]}$ is given by,

$$p'[t; A] = \sum_{k=0}^{n} k a_k t^{k-1} = a_1 + 2a_2 t + 3a_3 t^2 + \ldots + n a_n t^{n-1} \tag{5}$$

where the filter coefficients $A^* = \{a_1, \ldots, a_n\} \in \mathbb{R}$ are derived using least squares,

$$A^* = \arg\min_A (\xi_n) \quad \xi_n = \sum_{\tau=-m}^{m} \left( p[t + \tau; A] - x[t + \tau] \right)^2 = \sum_{\tau=-m}^{m} \left( \sum_{k=0}^{n} a_k t^k - x[t + \tau] \right)^2 \tag{6}$$

where $\tau \in [-m, m]$ is the "half width" of the approximation interval. The least-squares criterion requires minimizing the sum of squared error between $x[t]$ and $p[t;A]$ over the interval $[t−m, t+m]$. Here, the filter size $2m + 1$ and polynomial order $n$ determine the frequency response of the smoothed signal. To estimate cognitive load, it is crucial to compute the ratio between low and high-frequency pupil oscillations. Therefore, RIPA runs two filters in parallel with different $n, m$ values to capture low- and high-frequency oscillations, and computes their ratio. Next, it applies modulus maxima on the computed ratios, counts the values greater than threshold $\lambda$, and returns this as output (Jayawardena et al. 2022b). Thus, RIPA is proportional to pupil diameter oscillations, and hence, cognitive load.

Figure 2. Visualizations of gaze measures in the analytics dashboard. Left: Traditional positional gaze measures; Middle: Coefficient $\mathcal{K}$ measure; Right: RIPA Measure

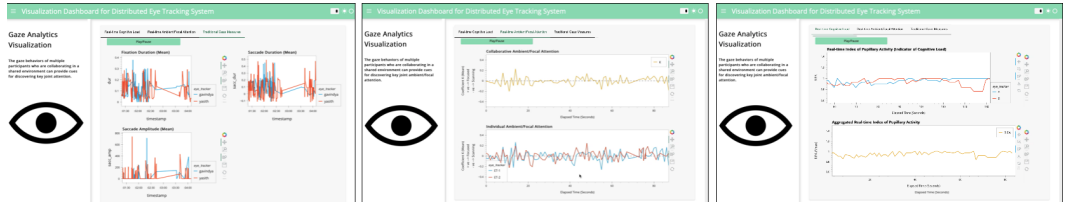

In our implementation, we use a non-overlapping sliding window of size *f*, where *f* is the sampling frequency of the signal.

For a multi-user environment, we extend RIPA by defining group RIPA as the average across all users, either within some time window or across the entire experiment. Our study uses *RIPA_window* for visualizations and *RIPA_experiment* when comparing group performance.

### 3.4 A-DisETrac Dashboard

A-DisETrac dashboard provides a detailed real-time visualization of (1) advanced gaze measures for each user ($^{w}\mathcal{K}$), the group ($^{Uw}\mathcal{K}$), RIPA, and group RIPA, and (2) traditional positional gaze measures for each user for the ongoing experiment (see Figure 2). Further, this dashboard provides more interactive functionalities to monitor, analyze, and control the gaze measure visualizations. A-DisETrac dashboard has four main key components as illustrated in Figure 3.

1. **Tabs:** Tabs allow the proctor to switch between the views of different gaze measure types. The views of different types of gaze measures that are designed in the dashboard (advanced gaze measures and traditional positional gaze measures) are shown in Figure 2.
2. **Play/Pause Control:** As the gaze measures are visualized in real-time charts (data streaming charts), they automatically update themselves after every *n* second. Hence, this play/pause control allows the proctor to pause the real-time charts and replay as necessary.
3. **Gaze Measures:** Real-time visualization of gaze measures calculated during the user experiment.
4. **Controls:** The control widgets include box zoom, wheel zoom, save, and reset.

### 3.5 User Studies

Next, we demonstrate the utility of the A-DisETrac dashboard by conducting two user studies. We evaluate their attention and cognitive load in real-time during **collaborative** and **competitive** tasks using advanced gaze and pupillary measures.

#### 3.5.1 Puzzle Solving Task

We conducted a pilot user study comprising ten participants (6M, 4F) and evaluated their attention and cognitive load in a collaborative activity. We conducted the study as physically isolated pairs (chosen randomly) collaborating online. The participants were graduate students in Computer Science and aged between 25-35 years. All the participants had normal or corrected-to-normal vision. We selected an online collaborative Jigsaw[3] puzzle-solving activity comprising a 50-piece jigsaw puzzle game (see Figure 4).

We used identical computer setups for each user comprising of desk-mounted GazePoint GP3[4] eye tracker, a 23.8-inch screen (1920x1080) (See Figure 5). The eye trackers operated at 60 Hz, and our setup sampled data at 30 Hz from the eye trackers. We hosted the MQTT broker and the analytics dashboard on another two computers connected through the public network. Considering that each session lasted less than 10 minutes, we synchronized all the devices only once at the beginning of each session.

**Figure 3. Layout of the A-DisETrac dashboard illustrating key components. 1. Tabs to switch between views, 2. Play and Pause the real-time charts, 3. Real-time visualization of gaze measures, 4. Control widgets to zoom, save, and reset.**

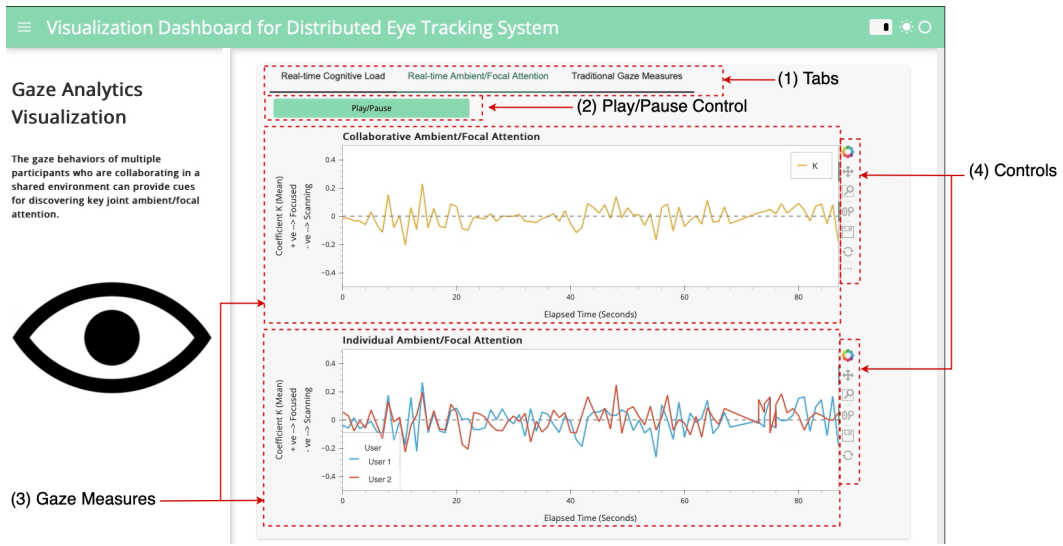

Each session started with a proctor calibrating each eye tracker using the standard 9-point calibration and manually testing the accuracy of the calibration. Then, the proctor presented a similar jigsaw puzzle as in the activity, explained the controls in the user interface, and allowed users to familiarize themselves with the activity. Meantime, we started the transmission, processing, and visualizations to ensure proper data flow. Once everything was in order, we presented the puzzle activity to the users and recorded the experimental data. During the experiment, we collected gaze location data and pupil data from the eye trackers. Further, we measured the time each pair took to complete the task.

We utilized the data restreaming setup to stream the collected gaze and pupil data. Upon reception at the A-DisETrac dashboard, we formed the coefficient $\mathcal{K}$ measure and RIPA values. When forming the windowed coefficient measure, we used a window of $w$ = 3000ms, sliding at each 300ms.

### 3.5.2 Battleship Game

In the second pilot study, we recruited ten volunteers (8 M, 2 F) and analyzed their attention and cognitive load during a **competitive** activity. Similar to the previous study, we conducted the study in physically isolated pairs, competing online. All the participants were graduate students in computer science and aged between 25-35 years with normal or corrected-to-normal vision. We used an online battleship[5] board game as the competitive activity, where each player strategically attempts to locate opponents battleships in a 10x10 square grid (see Figure 6).

Similar to the previous setup, we used two identical computer setups, each with a desk-mounted GazePoint GP3 eye tracker and a 27-inch screen (1920x1080) (See Figure 7). We sampled the data at 30 Hz and transferred the data to the analytics dashboard through an MQTT broker across a local network. Considering the duration of the experiment (< 5 mins), we synchronized only once during the start of the study session.

At the beginning of each session, a proctor calibrated the eye trackers using standard 9-point calibration and manually tested the accuracy. Then, the proctor briefed the experiment and started a practice session for the game. During the session, the proctor explained the game, controls, and scoring scheme while allowing the users to familiarize themselves with the activity. Once the participants were familiar and confident with the controls, we presented them with the competitive activity and initiated data collection procedures.

**Figure 4. An example of the online jigsaw puzzle solving activity**

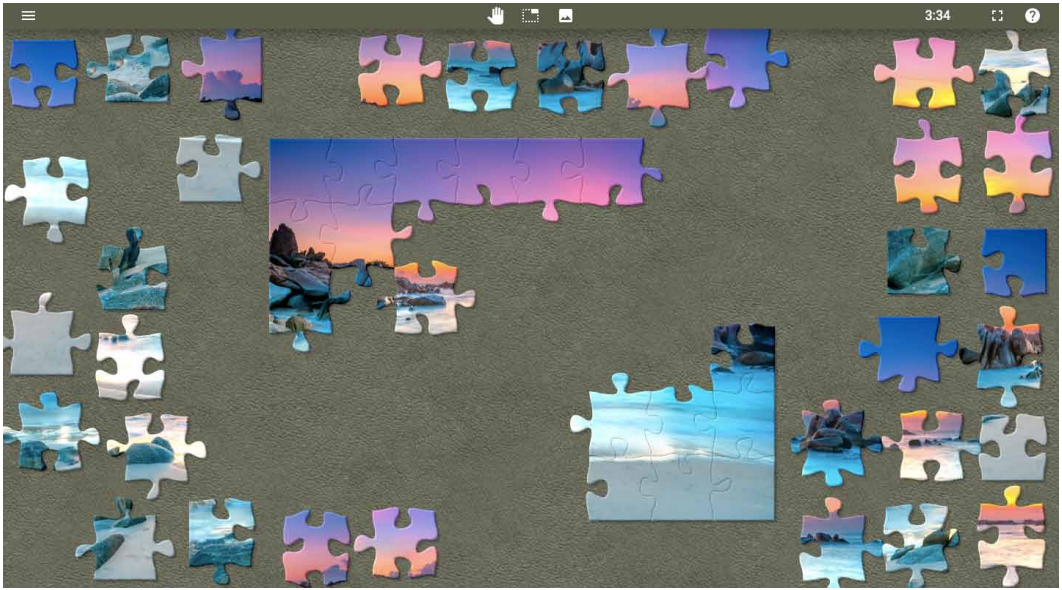

During the experiment, we collected gaze location data and pupil data from the eye trackers and formed advanced gaze measures upon reception at the A-DisETrac dashboard. Further, we collected the events on the battleship gameboard, such as success/failure strikes on enemy ships, through a browser extension.

### 3.5.3 Dashboard Evaluation

We conducted a user experience evaluation of our A-DisETrac dashboard with ten participants and compared it against DisETrac (Mahanama et al. 2023). The participants were graduate students (6M,

**Figure 5. Experimental setup of online jigsaw puzzle solving activity. A physically isolated pair of participants collaborating to solve an online jigsaw puzzle**

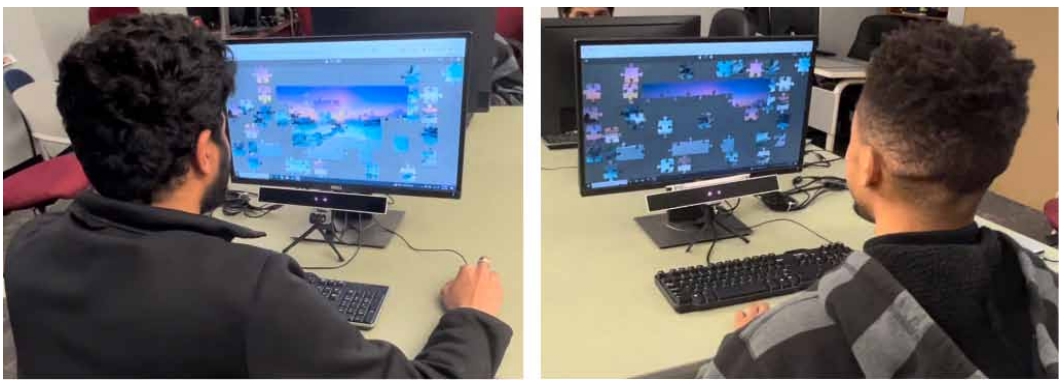

**Figure 6. An example of the online battleship game: The left is the opponent's board and the right is the board of the player**

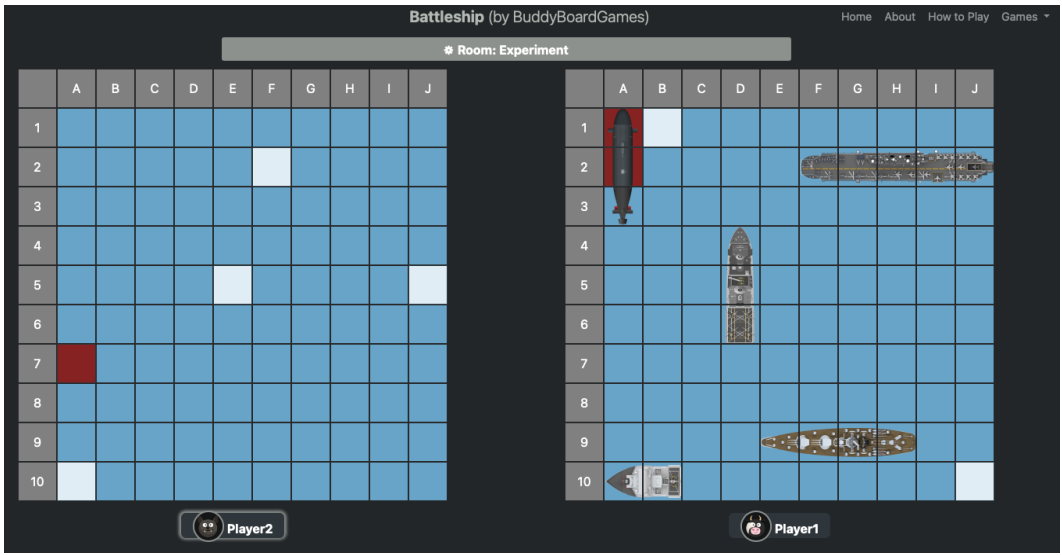

4F) in the Computer Science Department. For the evaluation, we used UEQ (Schrepp et al. 2019), a fast and reliable questionnaire to measure the User Experience of interactive products. The UEQ stands as a commonly utilized questionnaire designed to assess user experience across six scales: Attractiveness, Efficiency, Perspicuity, Dependability, Stimulation, and Novelty (See Table 7). Those constructs represent distinct attributes that help assess the acceptability of products. For each participant, we presented both dashboards with simulated data. Once the participants had used both dashboards, we provided them with the UEQ and asked them to provide feedback regarding their experience with each dashboard. To avoid the sequence effect, the two dashboards were presented in random order per participant.

**Figure 7. Experimental setup of online battleship game. A physically isolated pair of participants playing against each other in an online battleship game.**

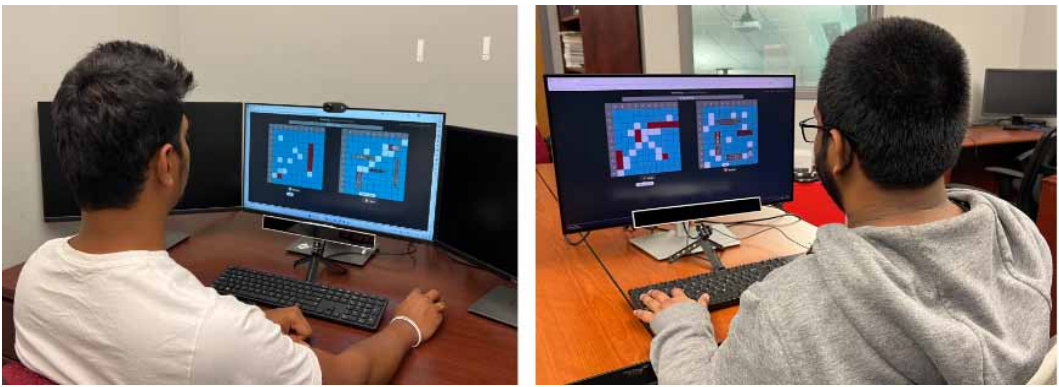

**Table 1. Data latency (gaze and pupil data) during the collaborative activity**

| Session | Mean Latency (ms) | Max Latency (ms) |
|---|---|---|
| 1 | 394 ± 235 | 973 |
| 2 | 408 ± 301 | 976 |
| 3 | 398 ± 313 | 1035 |
| 4 | 421 ± 330 | 1001 |
| 5 | 414 ± 341 | 1019 |
| Mean | 407 ± 308 | 994 |

## 4 RESULTS

### 4.1 Puzzle Solving Task

#### 4.1.1 Latency Analysis

We computed the latency by computing the delay between the transmission from the originating device to the destination dashboard in our system. We considered all eye-tracking data messages received during the experiment for the computation, assuming the effect of clock drifts to be negligible. Our results (see Table 1) indicate that our setup transmitted data with a mean latency of 407 ms and an average maximum latency of 994 ms in a public network. This indicates that our approach can notify a proctor of changes on average $^{w}\mathcal{K}$ in $d + 407$ ms, where $d$ represents the duration of the last fixation. To emulate potential real-world conditions, we did not adjust the quality of service parameters of the network to prioritize our data.

#### 4.1.2 Ambient/Focal Attention Analysis

To demonstrate the potential utility of the proposed windowed coefficient of attention, we investigated the relationship between group performance using the time to complete the puzzle and the coefficient of each group $^{UW}\mathcal{K}$. We observed all the groups to show negative $^{UW}\mathcal{K}$ values indicating ambient visual scanning behavior for all groups (see Table 2). Further, our investigation of the Pearson correlation between $^{UW}\mathcal{K}$ and time for completion revealed a strong negative correlation coefficient ($r = -0.9722$, $p = 0.0056$). This indicates that groups with more ambient attention (indicated by higher negative $^{UW}\mathcal{K}$) are associated with the group taking more time to complete the activity.

#### 4.1.3 RIPA Analysis

We analyzed the relationship between group performance using the time to complete the puzzle and the RIPA values of each group. We observed high RIPA values indicating increased cognitive load for all groups (see Table 3). We further analyzed the Pearson correlation between group RIPA and time for completion. We observed a moderate correlation coefficient ($r = 0.5804, p = 0.3049$).

### 4.2 Battleship Game

#### 4.2.1 Latency Analysis

Similar to the collaborative activity, we computed the latency in communication by using the timestamps of the originating device and the destination device, considering all the eye-tracking messages transmitted during the experiment while neglecting the effect of clock drifts. Compared to the previous study, we observed a lower mean latency during the experiment of 302 ms (see Table 4). Since we did not modify any quality of service parameters during the sessions, we attribute the lower latency to prevailing network conditions, such as traffic during the period of the experiment. This indicates the potential of improving the feedback time to a proctor through the quality of control parameters of the network.

**Table 2. Ambient/Focal Attention with Coefficient ($^{UW}\mathcal{K}$) during the experiment**

| Session | Attention Coefficient ($^{UW}\mathcal{K}$) | $\sigma$ | Total time (s) |
|---|---|---|---|
| 1 | -0.0515 | 0.4307 | 261 |
| 2 | -0.0350 | 0.4386 | 174 |
| 3 | -0.0375 | 0.4737 | 207 |
| 4 | -0.0350 | 0.3273 | 168 |
| 5 | -0.0996 | 0.4451 | 365 |

**Table 3. RIPA values of participant groups during the experiment**

| Session | RIPA | $\sigma$ | Total time (s) |
|---|---|---|---|
| 1 | 0.9505 | 0.0037 | 261 |
| 2 | 0.9588 | 0.0028 | 174 |
| 3 | 0.9478 | 0.0035 | 207 |
| 4 | 0.8153 | 0.2856 | 168 |
| 5 | 0.9795 | 0.0259 | 365 |

### 4.2.2 Ambient/Focal Attention Analysis

We demonstrated the utility of the proposed windowed coefficient of attention for a second pilot study. We compared the coefficient $^{W}\mathcal{K}$ values of participants for each session. We observed most of the participants have negative $^{UW}\mathcal{K}$ values indicating ambient visual scanning behavior (see Table 5). We did not observe a significant difference in coefficient $^{W}\mathcal{K}$ values between the winner and loser in each round. This indicates a similar visual scanning behavior between the players during a competitive gameplay task.

**Table 4. Data latency (gaze and pupil data) during the competitive activity**

| Session | Mean Latency (ms) | Max Latency (ms) |
|---|---|---|
| 1 | 679 ± 368 | 1324 |
| 2 | 608 ± 371 | 1279 |
| 3 | 170 ± 133 | 545 |
| 4 | 28 ± 29 | 327 |
| 5 | 23 ± 29 | 303 |
| Mean | 302 ± 186 | 755.6 |

Table 5. Coefficient $^W\mathcal{K}$ values of winner and loser of each the battleship rounds

| Session | Winner | Loser |
| --- | --- | --- |
| 1 | -0.031 | 0.024 |
| 2 | 0.0431 | -0.120 |
| 3 | -0.213 | -0.120 |
| 4 | -0.323 | -0.170 |
| 5 | -0.251 | -0.060 |

### 4.2.3 RIPA Analysis

We observed the winner of the majority of the sessions to associate with a higher average RIPA. However, the relationship did not yield any statistical significance, potentially indicating the competitive nature of the activity leading to higher cognitive load on both participants. We further investigated the relationship between the hit/miss of a strike during the game. Here, we investigated the point biserial coefficient used to measure the relationship when a variable is dichotomous-hit/miss on an enemy ship. As the measure of pupillary response, we used the increment in RIPA following the event with respect to preceding the event. Our test yielded no statistically significant relationship between the two variables within each session and across all sessions, which was indicated by a low correlation coefficient.

## 4.3 Dashboard Evaluation

We used the UEQ Data Analysis Tool, which uses T-Test (Kim 2015) with a 95% confidence interval to analyze the UEQ responses. The 26 items in the UEQ are categorized into six scales (see Table 7) that cover a comprehensive impression of user experience. We compared the scale means of the two dashboards as depicted in Figure 8. Our analysis did not show a statistically significant difference between the A-DisETrac dashboard and the DisETrac dashboard for the UEQ scales with $α = 0.05$ (see Table 8).

The UEQ tool offers a benchmark that helps to interpret the results and the benchmark relies on a number of studies concerning different products (Schrepp et al. 2019). We compared the results obtained for our A-DisETrac dashboard with the benchmark to gain insight into the user experience quality of our visualization dashboard compared to typical products in the market.

As illustrated in Figure 9, the A-DisETrac dashboard has "excellent" results in "Attractiveness" compared to the benchmark. Moreover, the A-DisETrac dashboard introduced in this work shows "good" results in "Efficiency", "Stimulation", and "Novelty" scales which is 75% better than the results in the benchmark data set. However, for the "Perspicuity" and "Dependability" scales, our A-DisETrac dashboard is better than 25% of the results in the benchmark which indicates as "below average".

Table 6. RIPA values of winner and loser of each the battleship rounds

| Session | Winner | Loser |
| --- | --- | --- |
| 1 | 0.96 | **0.97** |
| 2 | **0.97** | 0.96 |
| 3 | **0.97** | 0.96 |
| 4 | 0.95 | **0.97** |
| 5 | **0.96** | 0.83 |

**Table 7. Scales of user experience questionnaire**

| Scale | Definition |
|---|---|
| Attractiveness | Overall impression of the product. Do users like or dislike the product? |
| Perspicuity | Is it easy to get familiar with the product? Is it easy to learn how to use the product? |
| Efficiency | Can users solve their tasks without unnecessary effort? |
| Dependability | Does the user feel in control of the interaction? |
| Stimulation | Is it exciting and motivating to use the product? |
| Novelty | Is the product innovative and creative? Does the product catch the interest of users? |

**Figure 8. Comparison of scales means of the two dashboards. A-DisETrac Dashboard has higher values in "attractiveness," "efficiency," "dependability," "simulation," and "novelty" compared to our previous DisEtrac dashboard. DisEtrac dashboard has higher values in "Perspicuity." However, the results are not significant.**

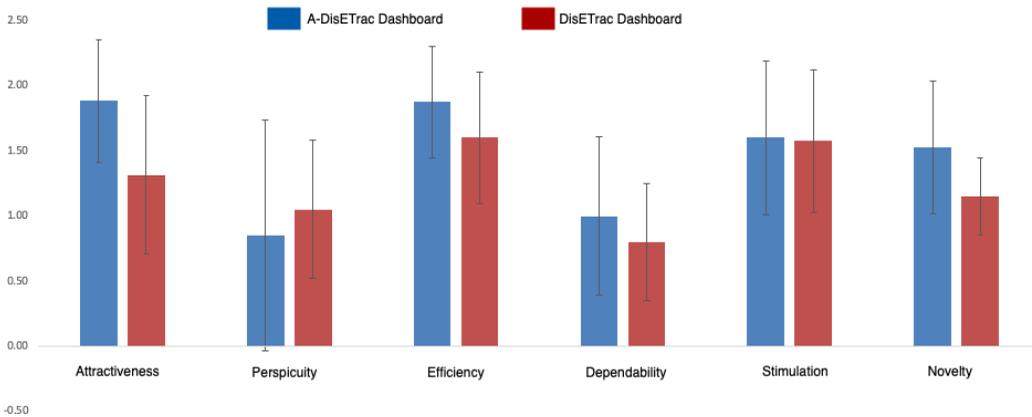

## 5 DISCUSSION

In this study, we implemented a real-time ambient/focal attention coefficient $\mathcal{K}$ and extended the concept to distributed multi-user eye tracking systems. Even though we demonstrate the utility through pilot studies, our approach requires further validation to determine the potential usage in analyzing user behaviors. Moreover, our study lacks an investigate on defining the ideal window size ($w$) and remains unexplored. However, our puzzle-solving user study revealed that the time a group takes to complete a puzzle is related to the ambient visual scanning behavior quantified by $^{UW}\mathcal{K}$. Our results indicate that groups that spent more time had more scanning of the screen and searching behavior. Considering that jigsaw puzzle solving requires the participants to identify and match pieces based on their visual characteristics (color, shape, texture), we presume a relationship exists between the ambient scanning behavior and the finding of a matching piece.

**Table 8. UEQ scale means and main effect of dashboards on UEQ scales**

| Scale | DisETrac | | Dashboard | | $p$ |
|---|---|---|---|---|---|
| | $\mu$ | $\sigma$ | $\mu$ | $\sigma$ | |
| Attractiveness | 1.32 | 0.98 | 1.88 | 0.76 | 0.1670 |
| Perspicuity | 1.05 | 0.86 | 0.85 | 1.43 | 0.7105 |
| Efficiency | 1.60 | 0.82 | 1.88 | 0.69 | 0.4273 |
| Dependability | 0.80 | 0.72 | 2.00 | 0.98 | 0.6104 |
| Stimulation | 1.58 | 0.88 | 1.60 | 0.95 | 0.9521 |
| Novelty | 1.15 | 0.47 | 1.53 | 0.82 | 0.2307 |

Figure 9. Comparison of scales means of A-DisETrac dashboard against the benchmark. Our dashboard has "excellent" results in "attractiveness," "good," results in "efficiency," "stimulation," "novelty," and "below average," in "perspicuity" and "dependability" compared to the benchmark.

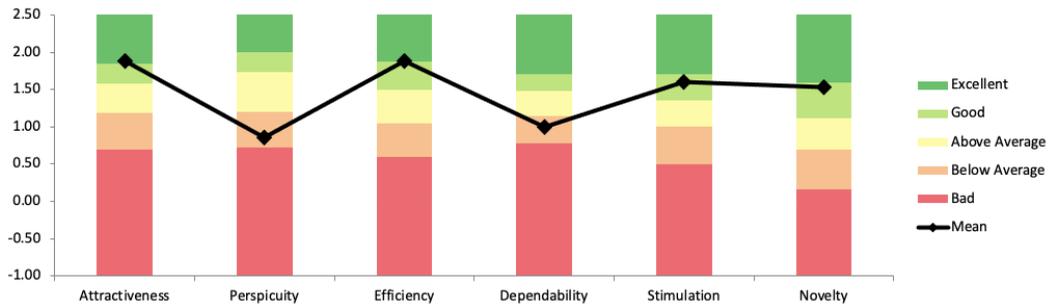

Moreover, we integrated RIPA into a distributed multi-user eye-tracking system and evaluated users' cognitive load during collaborative and competitive tasks. Our results showed a higher average RIPA indicating higher cognitive load in participants during both tasks. Battleship game requires players to manage information about potential ship placements and keep track of previous shots which can contribute to an increased cognitive load during gameplay. Similarly, in a jigsaw puzzle-solving activity, the players need to process visual information (the puzzle pieces), consider spatial relationships between pieces, and integrate individual pieces into the overall picture. This could potentially cause players to experience increased cognitive load. However, selecting the appropriate filter size ($2m + 1$) and polynomial order ($n$) is essential when defining Savitzky-Golay filters for calculating the ratio between low and high-frequency pupil oscillations in RIPA computations. Among the other limitations of our study, we did not optimize the hyperparameters in the RIPA algorithm specifically for our eye-tracker frequency. The RIPA algorithm we employed was originally tested with a 1000 Hz eye tracker frequency, while ours operates at 30 Hz. We can further fine-tune the hyperparameters in the RIPA algorithm to uncover subtle differences in cognitive load in battleship games and jigsaw puzzle-solving activities in the future.

A trivial approach to determine the effective and efficient means of computing variations of K and finding optimal parameters for the RIPA algorithm would be to conduct a comprehensive set of user studies encompassing different combinations of user behaviors. However, this approach could be costly and time-consuming. Alternatively, we can use synthetic data or restream data from previous experiments to investigate the broad spectrum of possibilities for variations of K and optimize the RIPA.

The results of the dashboard evaluation do not indicate a significant difference between the UEQ scales in our dashboard compared to the existing DisETrac dashboard (see Table 8). We further compared the UEQ results obtained for the A-DisETrac dashboard against the benchmark. The results indicate that the overall impression of our interactive dashboard is in the range of the 10% best results. However, the results of UEQ indicate that our dashboard is difficult to get familiar with and learn how to use compared to the average results in the benchmark. We believe having eye-tracking specific measures in our dashboard caused this low score as the majority of evaluators are not eye-tracking experts. We mainly focused on data visualization and analysis aspects in our dashboard rather than data security. Hence, we observed that the "Dependability" scale results of our dashboard are below the average of the benchmark. The "Dependability" is interpreted in the sense that the interaction is safe and controllable by the user. However, according to UEQ analysis, our dashboard has provided users with exciting and motivating experiences, allowed users to put in less effort, and caught users' interests compared to 75% of results in the benchmark data set.

To the best of our knowledge, no work exists in the literature for real-time visualization of advanced measures and system evaluation in multi-user eye-tracking capable of providing a fair comparison other than our previous work (Abeysinghe et al. 2023; Mahanama et al. 2023). Therefore, our study provides a benchmark for delineating limitations and advancements in the domain of advanced eye tracking measures in multi-user environments.

## 6 CONCLUSION

In this study, we presented a distributed eye-tracking system with real-time advanced gaze measures. Our setup uses off-the-shelf eye trackers connected through a public network for providing real-time insights on a multi-user eye-tracking experiment with advanced gaze measures. We presented the real-time gaze measures through an interactive dashboard. In the future, we plan to improve through the incorporation of Gaze Transition Entropy in multi-user distributed environments. Further, we plan to integrate real-time scan-path visualizations in our dashboard by streaming user viewports.

## ACKNOWLEDGEMENT

This work was supported in part by NSF 2045523. Any opinions, findings conclusions, or recommendations expressed in this material are the author(s) and do not necessarily reflect those of the sponsors.

# REFERENCES

Abeysinghe, Y. (2023). Gaze Analytics Dashboard for Distributed Eye Tracking. In: *2023 IEEE 24th International Conference on Information Reuse and Integration for Data Science (IRI)*. IEEE. doi:10.1109/IRI58017.2023.00031

Alharthi, S. A., Raptis, G. E., Katsini, C., Dolgov, I., Nacke, L. E., & Toups Dugas, P. O. (2021). Investigating the effects of individual cognitive styles on collaborative gameplay [TOCHI]. *ACM Transactions on Computer-Human Interaction*, *28*(4), 1–49. doi:10.1145/3445792

Alharthi, S. (2018). Toward understanding the effects of cognitive styles on collaboration in multiplayer games". In: *Companion of the 2018 ACM conference on computer supported cooperative work and social computing*, (pp. 169–172). ACM.

Ang, C. S., Zaphiris, P., & Mahmood, S. (2007). A model of cognitive loads in massively multiplayer online role playing games. *Interacting with Computers*, *19*(2), 167–179. doi:10.1016/j.intcom.2006.08.006

Berg, D. (2009). Free viewing of dynamic stimuli by humans and monkeys. In: *Journal of vision, 9,* 19–19.

Blascheck, T. (2017). Visualization of eye tracking data: A taxonomy and survey. In: *Computer Graphics Forum, 36*(8), 260–284. doi:10.1111/cgf.13079

Blattgerste, J., Renner, P., & Pfeiffer, T. (2018). Advantages of eye-gaze over headgaze-based selection in virtual and augmented reality under varying field of views. In: *Proceedings of the Workshop on Communication by Gaze Interaction*, (pp. 1–9). IEEE.

Brennan, S. (2008). Coordinating cognition: The costs and benefits of shared gaze during collaborative search. In: *Cognition, 106*(3), 1465–1477.

Cheng, S., Wang, J., Shen, X., Chen, Y., & Dey, A. (2022a). Collaborative eye tracking based code review through real-time shared gaze visualization. *Frontiers of Computer Science*, *16*(3), 1–11. doi:10.1007/s11704-020-0422-1

Cheng, S., Wang, J., Shen, X., Chen, Y., & Dey, A. (2022b). Collaborative eye tracking based code review through real-time shared gaze visualization. *Frontiers of Computer Science*, *16*(3), 163704. doi:10.1007/s11704-020-0422-1

Convertino, G. (2003). Exploring context switching and cognition in dual-view coordinated visualizations. In: *Proceedings International Conference on Coordinated and Multiple Views in Exploratory Visualization-CMV 2003*. IEEE. doi:10.1109/CMV.2003.1215003

Cotton, K. (2023). *The Effects of Mind-Wandering, Cognitive Load and Task Engagement on Working Memory Performance in Remote Online Experiments*. Research Gate.

D'Angelo, S., & Begel, A. (2017). Improving communication between pair programmers using shared gaze awareness. In: *Proceedings of the 2017 CHI conference on human factors in computing systems*, (pp. 6245–6290). ACM. doi:10.1145/3025453.3025573

D'Angelo, S., & Gergle, D. (2016a). Gazed and confused: Understanding and designing shared gaze for remote collaboration. In: *Proceedings of the 2016 CHI Conference on Human Factors in Computing Systems*, (pp. 2492–2496). ACM. doi:10.1145/2858036.2858499

D'Angelo, S., & Gergle, D. (2016b). Gazed and confused: Understanding and designing shared gaze for remote collaboration. In: *Proceedings of the 2016 chi conference on human factors in computing systems*, (pp. 2492–2496). ACM. doi:10.1145/2858036.2858499

D'Angelo, S., & Gergle, D. (2018). An eye for design: gaze visualizations for remote collaborative work. In: *Proceedings of the 2018 CHI Conference on Human Factors in Computing Systems*, (pp. 1–12). ACM. doi:10.1145/3173574.3173923

D'Angelo, S., & Schneider, B. (2021). Shared gaze visualizations in collaborative interactions: Past, present and future. *Interacting with Computers*, *33*(2), 115–133. doi:10.1093/iwcomp/iwab015

Drusch, G. (2014). Analysing eye-tracking data: From scanpaths and heatmaps to the dynamic visualisation of areas of interest. In: Advances in science, technology, higher education and society in the conceptual age: STHESCA, 20(205), 25.

Duchowski, A. T. (2018). The Index of Pupillary Activity: Measuring Cognitive Load Vis-à-Vis Task Difficulty with Pupil Oscillation. In: *Proceedings of the 2018 CHI Conference on Human Factors in Computing Systems*. Montreal QC, Canada: Association for Computing Machinery. doi:10.1145/3173574.3173856

Duchowski, A. T. (2020). The Low/High Index of Pupillary Activity. In: *Proceedings of the 2020 CHI Conference on Human Factors in Computing Systems*. Association for Computing Machinery. doi:10.1145/3313831.3376394

Duchowski, A. (2012). Aggregate gaze visualization with real-time heatmaps. *Proceedings of the symposium on eye tracking research and applications*, (pp. 13–20). ACM.

Garcia, A. (2003). A distributed real time eye-gaze tracking system. In: *EFTA 2003. 2003 IEEE Conference on Emerging Technologies and Factory Automation. Proceedings (Cat. No. 03TH8696)*. IEEE. doi:10.1109/ETFA.2003.1248746

Guo, J., & Feng, G. (2013). How eye gaze feedback changes parent-child joint attention in shared storybook reading? an eye-tracking intervention study. In: Eye gaze in intelligent user interfaces: Gaze-based analyses, models and applications. ACM.

Jayawardana, Y. (2022). StreamingHub: interactive stream analysis workflows. In: *Proceedings of the 22nd ACM/IEEE Joint Conference on Digital Libraries*, (pp. 1–10). IEEE.

Jayawardena, G. (2022a). Introducing a Real-Time Advanced Eye Movements Analysis Pipeline. In: *2022 Symposium on Eye Tracking Research and Applications*, (pp. 1–2). ACM. doi:10.1145/3517031.3532196

Jayawardena, G. (2020b). "Pilot study of audiovisual speech-in-noise (sin) performance of young adults with adhd. In: *Proceedings of the Symposium on Eye Tracking Research and Applications*. ETRA. New York, NY, USA: Association for Computing Machinery. doi:10.1145/3379156.3391373

Jayawardena, G., & Jayarathna, S. (2021a). Automated Filtering of Eye Movements Using Dynamic AOI in Multiple Granularity Levels [IJMDEM]. *International Journal of Multimedia Data Engineering and Management*, *12*(1), 49–64. doi:10.4018/IJMDEM.2021010104

Jayawardena, G., Jayawardana, Y., Jayarathna, S., Högström, J., Papa, T., Akkil, D., Duchowski, A. T., Peysakhovich, V., Krejtz, I., Gehrer, N., & Krejtz, K. (2022b). Toward a Real-Time Index of Pupillary Activity as an Indicator of Cognitive Load. *Procedia Computer Science*, *207*, 1331–1340. doi:10.1016/j.procs.2022.09.189

Johnson, D. (2015). Cooperative game play with avatars and agents: Differences in brain activity and the experience of play. In: *Proceedings of the 33rd annual acm conference on human factors in computing systems*, (pp. 3721–3730). ACM. doi:10.1145/2702123.2702468

Kim, T. K. (2015). T test as a parametric statistic. *Korean Journal of Anesthesiology*, *68*(6), 540–546. doi:10.4097/kjae.2015.68.6.540 PMID:26634076

Krejtz, K. (2014). Entropy-based statistical analysis of eye movement transitions. In: *Proceedings of the Symposium on Eye Tracking Research and Applications*. New York, NY, USA: Association for Computing Machinery. doi:10.1145/2578153.2578176

Krejtz, K. (2016). Discerning Ambient/Focal Attention with Coefficient K. In: ACM Transactions on Applied Perception. ACM. doi:10.1145/2896452

Krejtz, K. (2018). Eye tracking cognitive load using pupil diameter and microsaccades with fixed gaze. In: PloS one. doi:10.1371/journal.pone.0203629

Krejtz, K., Duchowski, A., Szmidt, T., Krejtz, I., González Perilli, F., Pires, A., Vilaro, A., & Villalobos, N. (2015). Gaze Transition Entropy. *ACM Transactions on Applied Perception*, *13*(1), 1–20. doi:10.1145/2834121

Kütt, G. H. (2019). Eye-write: Gaze sharing for collaborative writing. In: *Proceedings of the 2019 CHI Conference on Human Factors in Computing Systems*, (pp. 1–12). ACM. doi:10.1145/3290605.3300727

Langner, M., Toreini, P., & Maedche, A. (2022). EyeMeet: A Joint Attention Support System for Remote Meetings. In: *CHI Conference on Human Factors in Computing Systems Extended Abstracts*, (pp. 1–7). ACM. doi:10.1145/3491101.3519792

Mahanama, B. (2022a). Multi-User Eye-Tracking. In: *2022 Symposium on Eye Tracking Research and Applications*. ACM.

Mahanama, B. (2022b). Eye movement and pupil measures: A review. In: Frontiers in Computer Science, 3. doi:10.3389/fcomp.2021.733531

Mahanama, B. (2023). DisETrac: Distributed Eye-Tracking for Online Collaboration. In: *Proceedings of the 2023 Conference on Human Information Interaction and Retrieval*, (pp. 427–431). ACM. doi:10.1145/3576840.3578292

Mahanama, B., Jayawardena, G., & Jayarathna, S. (2021). Analyzing Unconstrained Reading Patterns of Digital Documents Using Eye Tracking. In: *2021 ACM/IEEE Joint Conference on Digital Libraries (JCDL)*. IEEE. doi:10.1109/JCDL52503.2021.00036

Mahanama, B. (2022c). *Multidisciplinary Reading Patterns of Digital Documents*. Research Gate.

Michalek, A. (2019). Predicting ADHD using eye gaze metrics indexing working memory capacity. In: *Computational Models for Biomedical Reasoning and Problem Solving*. IGI Global.

Neider, M. (2010). Coordinating spatial referencing using shared gaze. In: *Psychonomic bulletin & review, 17*, 718–724.

Pathirana, P., Senarath, S., Meedeniya, D., & Jayarathna, S. (2022). Eye gaze estimation: A survey on deep learning-based approaches. *Expert Systems with Applications*, *199*, 116894. doi:10.1016/j.eswa.2022.116894

Savitzky, A., & Golay, M. J. E. (1964). Smoothing and Differentiation of Data by Simplified Least Squares Procedures. In: *Analytical Chemistry*, *38,* 1627–1639. pubs.acs.org/doi/abs/10.1021/ac60214a047

Schneider, B., Sharma, K., Cuendet, S., Zufferey, G., Dillenbourg, P., & Pea, R. (2018). Leveraging mobile eye-trackers to capture joint visual attention in co-located collaborative learning groups. *International Journal of Computer-Supported Collaborative Learning*, *13*(3), 241–261. doi:10.1007/s11412-018-9281-2

Schrepp, M., & Thomaschewski, J. (2019). *Handbook for the modular extension of the User Experience Questionnaire*. UEQ. www. ueq-online. org

Senarath, S., Pathirana, P., Meedeniya, D., & Jayarathna, S. (2022). Customer gaze estimation in retail using deep learning. *IEEE Access : Practical Innovations, Open Solutions*, *10*, 64904–64919. doi:10.1109/ACCESS.2022.3183357

Sharma, K. Patrick Jermann, and Pierre Dillenbourg (2014). ""With-me-ness": A gaze-measure for students' attention in MOOCs". In: *Proceedings of international conference of the learning sciences 2014*. CONF. ISLS, pp. 1017–1022.

Sharma, K. (2016). *Visual augmentation of deictic gestures in mooc videos*. International Society of the Learning Sciences.

Smith, Tim J and Parag K Mital (2013). "Attentional synchrony and the influence of viewing task on gaze behavior in static and dynamic scenes". In: *Journal of vision* 13.8, pp. 16–16.

Špakov, O. (2019). "Eye gaze and head gaze in collaborative games". In: *Proceedings of the 11th ACM Symposium on Eye Tracking Research & Applications*, pp. 1–9.

Špakov, Oleg and Darius Miniotas (2007). "Visualization of eye gaze data using heat maps". In.

Stein, R. (2004). Another person's eye gaze as a cue in solving programming problems. In: *Proceedings of the 6th international conference on Multimodal interfaces*, (pp. 9–15). ACM. doi:10.1145/1027933.1027936

Wikström, V., Saarikivi, K., Falcon, M., Makkonen, T., Martikainen, S., Putkinen, V., Cowley, B. U., & Tervaniemi, M. (2022). Inter-brain synchronization occurs without physical co-presence during cooperative online gaming. *Neuropsychologia*, *174*, 108316. doi:10.1016/j.neuropsychologia.2022.108316 PMID:35810882

Zhang, Y., Pfeuffer, K., Chong, M. K., Alexander, J., Bulling, A., & Gellersen, H. (2017). Look together: using gaze for assisting co-located collaborative search. *Personal and Ubiquitous Computing*, *21*(1), 173–186. doi:10.1007/s00779-016-0969-x

Zhao, S., Cheng, S., & Zhu, C. (2023). 3D Gaze Vis: Sharing Eye Tracking Data Visualization for Collaborative Work in VR Environment. arXiv preprint arXiv.

**ENDNOTES**

1. https://developer.tobii.com/xr/learn/analytics/fundamentals/visualizations/
2. https://www.tableau.com/resource/eye-tracking-study
3. https://www.jigsawexplorer.com/
4. https://www.gazept.com/product-category/gp3/
5. https://buddyboardgames.com/battleship

*Yasasi Abeysinghe is a PhD student at Old Dominion University, where she is associated with the Web Science and Digital Libraries (WS-DL) research group. She is presently pursuing the areas of her research interest in eye tracking, human-computer interaction, and data science. Her contribution to these fields includes publishing papers in venues such as IEEE IRI, and ACM ETRA and reviewing papers for ACM CIKM, IEEE Big Data, ACM CHIIR, and ACM/IEEE JCDL conferences. Yasasi is the publicity chair of the ACM ETRA 2024 organizing committee. She graduated with a B.Sc. Engineering (Hons) degree in Computer Science and Engineering at the University of Moratuwa, Sri Lanka, with first-class honors, in December 2018.*

*Bhanuka Mahanama is a Ph.D. student at Old Dominion University working with the Neuro Information Retrieval and Data Science Lab (NirdsLab) and the Web Science and Digital Libraries (WS-DL) research group. His research interests include eye tracking, gaze estimation, multi-user environments, and data science. He has received the best poster award at ACM/IEEE JCDL 2021 and has published in venues such as ACM CHIIR, ETRA, Augmented Humans, and IEEE IRI. He is a member of the program committee at the ACM/IEEE JCDL and ACM CIKM and also reviews papers for ACM CHIIR, IUI, Augmented Human, and ETRA.*

*Gavindya Jayawardena is a Ph.D. student in Computer Science at Old Dominion University. She joined the program in 2019 and currently pursues research in the field of eye-tracking. Gavindya has contributed to this domain by publishing insightful papers on eye-tracking research. Her research interests span across eye-tracking, human-computer interaction, data science, and machine learning. Before commencing her Ph.D., she earned her B.Sc. degree in Computer Science and Engineering from University of Moratuwa, Sri Lanka, in 2018.*

*Yasith Jayawardana is a Ph.D. student in Computer Science at Old Dominion University. He joined the program in 2019 and conducts research in real-time analytics and deep learning. Yasith has published several papers on bio-signal analysis, real-time analytics, distributed systems, and deep learning. Before starting his Ph.D., he obtained his B.Sc. in Computer Science and Engineering from University of Moratuwa, Sri Lanka, in 2018.*

*Mohan Krishna is a dedicated PhD candidate with a strong background in data analysis and deep learning. He is currently pursuing a doctorate in computer science at Old Dominion University in Norfolk, Virginia. As a dedicated member of the Web Science and Digital Libraries (WS-DL) research group, he is committed to developing accessible and usable technology for visually impaired individuals. His research spans natural language processing (NLP), computer vision, human-computer interaction, accessibility, usability, machine learning, deep learning, probability and statistics, and generative AI. Specializing in designing, developing, deploying, and optimizing machine learning and deep learning models, Mohan's work has led to valuable contributions, including patents and publications in renowned conferences and journals such as EICS, IUI, CHIIR, Hypertext, JCDL, IRI, MDPI, and more. Mohan's academic journey includes a Bachelor of Technology in Information Technology and a Master of Technology in Computer Science and Engineering from Manipal Institute of Technology. During his studies, he interned at HP Hewlett-Packard Research and Development, where he focused on enhancing HP printer intelligence. He also had the privilege of completing a digital internship at P&G Procter & Gamble near Prague, Czech Republic, Europe, where he worked on IoT, data analysis, and web development, gaining valuable insights into the European market and its vibrant culture. Passionate about improving accessibility for the visually impaired through his expertise in computer vision and machine learning, Mohan is also an enthusiast of horror mystery filmmaking and dedicates his time to social and spiritual groups.*

*Andrew T. Duchowski is a professor of Computer Science at Clemson University. He received his B.Sc. ('90) and Ph.D. ('97) degrees in Computer Science from Simon Fraser University, Burnaby, Canada, and Texas A&M University, College Station, TX, respectively. His research and teaching interests include visual attention and perception, eye movements, computer vision, graphics, and virtual environments. He joined the School of Computing faculty at Clemson in January, 1998. He has since published a corpus of papers and a textbook related to eye tracking research, and has delivered courses and seminars on the subject at international conferences. He maintains Clemson's eye tracking laboratory, and teaches a regular course on eye tracking methodology attracting students from a variety of disciplines across campus.*

*Vikas Ashok is an Assistant Professor in the Department of Computer Science at Old Dominion University. Before joining Old Dominion University, he was a post-doctoral associate at Stony Brook University, where he graduated with a Ph.D. in Computer Science. His primary research focus is on Accessible Computing, and he is particularly interested in designing and developing personalized semantics-aware assistive technologies for people with vision impairments.*

*Sampath Jayarathna is an Assistant Professor of Computer Science at Old Dominion University, where he is associated with the Web Science and Digital Libraries (WS-DL) research group. Before joining ODU in 2018, he was an Assistant Professor at California State Polytechnic University Pomona. Dr. Jayarathna earned a Ph.D. in Computer science from the Texas A&M University College Station in 2016. His research interests include machine learning, information retrieval, data science, eye tracking, and brain-computer interfacing.*